\documentclass[aps,prl,floatfix,nofootinbib,tightenlines,twocolumn,amsmath,amssymb,footinbib]{revtex4-1}
\usepackage{hyperref}

\newcommand{\bra}[1]{\ensuremath{\left\langle #1\right\vert}}
\newcommand{\ket}[1]{\ensuremath{\left\vert #1\right\rangle}}

\newcommand{\hsp}[1]{\hspace{#1 em}}
\newcommand{\sqz}{\hsp{-0.1}}
\newcommand{\ketbra}[2]{\left\vert{#1}\right\rangle \sqz\sqz\sqz \left\langle{#2}\right\vert}

\begin{document}
\title{Extracting entanglement from identical particles}
\date{\today}

\author{N. Killoran, M. Cramer, and M. B. Plenio}
\affiliation{Institut f\"{u}r Theoretische Physik, Albert-Einstein-Allee 11, Universit\"{a}t Ulm, D-89069 Ulm, Germany}

\begin{abstract}
Identical particles and entanglement are both fundamental components of quantum mechanics. However, when identical particles are condensed in a single spatial mode, the standard notions of entanglement, based on clearly identifiable subsystems, break down. This has led many to conclude that such systems have limited value for quantum information tasks, compared to distinguishable particle systems. To the contrary, we show that any entanglement formally appearing amongst the identical particles, including entanglement due purely to symmetrization, can be extracted into an entangled state of independent modes, which can then be applied to any task. In fact, the entanglement of the mode system is in one-to-one correspondence with the entanglement between the inaccessible identical particles. This settles the long-standing debate about the resource capabilities of such states, in particular spin-squeezed states of Bose-Einstein condensates, while also revealing a new perspective on how and when entanglement is 
generated in passive optical networks. Our results thus reveal new fundamental connections between entanglement, squeezing, and indistinguishability.
\end{abstract}

\maketitle


Identical particles are essentially axiomatic in quantum mechanics \cite{feynmann10a}. Entanglement is another fundamental quantum concept, which serves, within the local operations \& classical communication (LOCC) paradigm, as a valuable resource for quantum information tasks \cite{plenio07a}. However, the notion of entanglement for identical particles is troublesome because one can have subsystems which cannot be operationally distinguished. For instance, what is a meaningful definition of entanglement when identical particles occupy the same spatial mode? How can we make sense of entanglement between subsystems that are not operationally accessible? 

Experimental progress in Bose-Einstein condensates (BECs), related to multiparticle entanglement, has magnified the existing debate \cite{esteve08a,riedel10a,gross10a,sorensen01a,sorensen01b,omar02a,micheli03a,hines03a, ghirardi04a,korbicz05a,cavalcanti07a, hyllus10b,hyllus12a,benatti11a,he11a,dalton13a,dalton14a}. In such experiments, spin-squeezed states \cite{kitagawa93a}, useful in high-precision metrology \cite{wineland94a,huelga97a}, are generated. Taking individual particles as subsystems, such states are highly entangled. In fact, due to symmetrization, all correlated states of identical particles are strongly multiparticle entangled \cite{ichikawa08a,wei10a}. But since our access to the designated subsystems is fundamentally restricted by indistinguishability, what is the use of this entanglement? Many authors share the viewpoint that such entanglement is a mathematical artifact, and not fully legitimate \cite{schliemann01a,schliemann01b,eckert02a,ghirardi77a,ghirardi02a,ghirardi04a,paskauskas01a,
zanardi02a,shi03a,
barnum04a,barnum05a,levay05a,esteve08a,plastino09a,benatti11a,tichy13a}. Entanglement that results purely from symmetrization has variously been described as unphysical \cite{zanardi02a, barnum04a}, inaccessible \cite{eckert02a,esteve08a}, and not a resource (in the standard quantum information sense) \cite{eckert02a,ghirardi04a,paskauskas01a,barnum04a,barnum05a,levay05a,esteve08a,plastino09a}. One can avoid this illusion by modifying the definition of entangled states \cite{schliemann01a,schliemann01b,eckert02a,ghirardi77a,ghirardi02a,ghirardi04a,paskauskas01a,shi03a,plastino09a}, or by defining entanglement relative to the subsystem structure of observables, not states \cite{barnum04a,barnum05a,zanardi02a,zanardi04a,sasaki11a,balachandran13a}. While correlated states 
of identical particles may indeed be useful for metrology \cite{pezze09a} \footnote{Note that for metrology applications, it is not essential to have access to the individual particles.}, the notion of multiparticle entanglement in such systems is seemingly flawed. 

In this Letter, we re-examine systems of indistinguishable particles, resurrecting legitimate meaning for their entanglement structure. Using intuition similar to \cite{wiseman03a}, entanglement should be given meaning only when it can be extracted onto distinguishable registers via operations which themselves do not contribute any entanglement. Such entanglement can then be applied to standard quantum information tasks. Remarkably, we show that this extractable entanglement exactly corresponds with the entanglement one would find within a naive multiparticle description. Specifically, identical particle entanglement can be transferred, with unit probability, onto independent modes using elementary operations. Thus, symmetrization entanglement is a fundamental, ubiquitous, and readily-extractable resource for standard quantum information tasks. Our results demonstrate the usefulness of single-mode BECs for many tasks beyond metrology, and reveal new insight on how and when entanglement is generated in 
passive optical networks.


Suppose we have $N$ identical particles in the same spatial mode. We focus on BECs, because they are the largest source of debate in the literature, but these ideas also apply to other scenarios, such as photons with polarization degrees of freedom (see \cite{cavalcanti07a} for a related fermionic example). Let each particle have two internal states (for convenience, called ``spin down/up"), so that individual particles have a two-dimensional Hilbert space $\mathcal{H}=\mathrm{span}\{\ket{0},\ket{1}\}$. Formally, we can describe the state of the $N$-particle system within the Hilbert space $\mathcal{H}_N := \mathcal{H}^{\otimes N}$. Implicit here is a specific pseudo-labeling of the particles: particle $p$ is associated with the $p$th Hilbert space in the decomposition. For identical particles, these pseudo-labels cannot be distinguished experimentally and have ambiguous physical meaning. 

Mathematically, however, this `first quantization' basis gives sufficient structure to consider entanglement. For $N>1$, we can imagine partitioning this space into disjoint subsystems containing fixed numbers of particles.  In the decomposition $\mathcal{H}_N=\mathcal{H}^{\otimes N}$, each particle is its own subsystem. On the other extreme, we can consider a bipartition, grouping the first $N_X$ particles into one subsystem and the remaining $N_Y$ into another, giving $\mathcal{H}_N = \mathcal{H}^{\otimes N_X}\otimes \mathcal{H}^{\otimes N_Y}=:\mathcal{H}_X\otimes\mathcal{H}_Y$. Here we will focus mainly on bipartite entanglement; extensions to multipartite scenarios are analogous (e.g., see Appendix). Alternatively, we can use a `second quantization' basis that more accurately describes the accessible degrees of freedom. For $N$ identical particles in mode $\mathbf{A}$, the symmetric states $\{\ket{n,N-n}\}$ enumerate composite states that have $n$ spin-down and $N-n$ spin-up particles. 
These can be 
obtained by symmetrizing single-particle states:
\begin{align}\label{eq:symmstate}
	\ket{n,N-n}_\mathbf{A} = \frac{1}{\sqrt{\binom{N}{n}}}\mathcal{S}\left[\ket{0}_1\dots\ket{0}_n\ket{1}_{n+1}\dots\ket{1}_N\right],
\end{align}
where $\mathcal{S}$ generates a sum over all unique permutations with $n$ spin-down particles out of $N$ and $\binom{N}{n}$ is the normalization. These states form an orthonormal basis for the symmetric subspace on which all physical states live. The symmetric subspace can also be generated using creation operators: $\hat{a}_0^\dagger\ket{k,l}_\mathbf{A} = \sqrt{k+1}\ket{k+1,l}_\mathbf{A}$ and $\hat{a}_1^\dagger\ket{k,l}_\mathbf{A} = \sqrt{l+1}\ket{k,l+1}_\mathbf{A}$. 


\paragraph{Mode-splitting.}

In the typical setting, bipartite entanglement is defined relative to two parties with independent, accessible subsystems. In contrast, in the full $N$-particle state space, the subsystems which appear to be entangled are inherently inaccessible. Intuitively, we might imagine getting at this entanglement by somehow splitting the particles up into physically distinguishable modes. For instance, we could let a BEC cloud spread until it separates into distinct clusters, or we could use a more tunable operation such as tunnelling into neighbouring modes. The occupied output modes then provide some physically accessible degrees of freedom, and we can safely speak of entanglement between these modes. 

But there are a few obvious concerns. First, we will still not have any access to the particle pseudo-labels that characterize the original state's entanglement. If we find that there is a single particle in output mode $\mathbf{C}$, we can use this information to distinguish this particle in future experiments. But relative to the original pseudo-labelling, ``the particle in mode $\mathbf{C}$'' remains some symmetric superposition of all identical particles from the initial state. So although one can consider mode entanglement in the output state, how is this related to the entanglement defined relative to the pseudo-labels?  The second issue is the mode-splitting process itself. Since we start with one mode and end with more than one, we have essentially performed a non-local operation. How do we know that entanglement between the output modes was not created by the splitting operation itself? Finally, for massive particles, there is the issue of superselection rules \cite{wiseman03a}, 
whereby superpositions of local particle numbers cannot be measured. How might this affect the entanglement we can extract?

To explore these issues, we consider the example of a beamsplitter transformation from optics. For BECs, this is equivalent to a tunnelling operation where particles can leak from mode $\mathbf{A}$ into a neighbouring mode $\mathbf{B}$ via a Hamiltonian of the form $H\sim \sum_{k=0,1}(\hat{b}_k^\dagger\hat{a}_k + \hat{a}_k^\dagger\hat{b}_k)$. We denote the modes post-tunnelling by $\mathbf{C}$ and $\mathbf{D}$. Suppose we initially have the 3-particle state $\ket{\phi_\mathrm{in}}_\mathbf{A}=\ket{2,1}_\mathbf{A}$, a symmetric state with 2 spin down and 1 spin up particles. Because of symmetrization, this state is entangled in the pseudo-label basis (for any non-trivial bipartition). We then apply a splitting transformation $\hat{a}_k^\dagger\rightarrow r\hat{c}_k^\dagger + t\hat{d}_k^\dagger$ ($k=0,1$); this operation, which is insensitive to the internal degrees of freedom, transfers single particles from mode $\mathbf{A}$ into $\mathbf{C}~(\mathbf{D})$ with amplitude $r~(t)$. The 
other mode $\mathbf{B}$ initially has no particles. The final state is
\begin{align}\label{eq:bsoutputstate}
	\ket{\phi_\mathrm{out}}_\mathbf{CD} =~& r^3\ket{2,1}_\mathbf{C}\ket{0,0}_\mathbf{D} \nonumber\\
					+ & \sqrt{3}r^2t\frac{1}{\sqrt{3}}\left[\ket{2,0}_\mathbf{C}\ket{0,1}_\mathbf{D} + \sqrt{2}\ket{1,1}_\mathbf{C}\ket{1,0}_\mathbf{D}\right]  \nonumber\\
					+ & \sqrt{3}rt^2\frac{1}{\sqrt{3}}\left[\ket{0,1}_\mathbf{C}\ket{2,0}_\mathbf{D} + \sqrt{2}\ket{1,0}_\mathbf{C}\ket{1,1}_\mathbf{D}\right]  \nonumber\\
					+ & t^3\ket{0,0}_\mathbf{C}\ket{2,1}_\mathbf{D}.
\end{align}
We have ordered the output state with respect to different possibilities for local particle numbers. In the first/last case (all particles in one mode), the output state is the same as the input state, with no mode entanglement. For the other cases, there is clearly entanglement between the output modes. Even if we project onto fixed local particle numbers to respect superselection rules \cite{wiseman03a}, we have, on average, non-zero entanglement in the output state. This entanglement is now a valid resource in the LOCC paradigm. Evidently, correlated single-mode states have some many-body coherence properties \cite{tichy13a} that may lead to mode entanglement after splitting. In fact, we recognize a conceptual connection with the widely-known notion from continuous-variable optics (\cite{kim02a,wolf03a,asboth05a,jiang13a}) that beamsplitters transform non-classical states (e.g., squeezed states) into mode entangled states.

But how does this output mode entanglement relate to the input state's apparent pseudo-label entanglement? For concreteness, suppose we group particles 1 and 2 into subsystem $X$ and particle 3 into subsystem $Y$. To classify the entanglement, we put $\ket{\phi_\mathrm{in}}_\mathbf{A}$ into Schmidt form: 
\begin{align}\label{eq:bsinputstate}
	\ket{2,1}_\mathbf{A} = & \frac{1}{\sqrt{3}}\big( \ket{0}_1 \ket{0}_2 \ket{1}_3 + \ket{0}_1 \ket{1}_2 \ket{0}_3 + \ket{1}_1 \ket{0}_2\ket{0}_3 \big)\nonumber\\
	 = & \frac{1}{\sqrt{3}}\left( \left[ \ket{0}_1\ket{0}_2 \right] \ket{1}_3 + \sqrt{2}\left[\frac{1}{\sqrt{2}}\mathcal{S}[\ket{0}_1\ket{1}_2]\right] \ket{0}_3 \right)\nonumber\\
	 = & \frac{1}{\sqrt{3}}\left( \ket{2,0}_{X} \ket{0,1}_Y + \sqrt{2}\ket{1,1}_{X} \ket{1,0}_Y \right).
\end{align}
In the last line we have rewritten the states within the fictitious subsystems $X$ and $Y$ in second-quantized form. We now make the crucial observation that Eq. (\ref{eq:bsinputstate}) is algebraically equivalent to the mode-split state in Eq. (\ref{eq:bsoutputstate}) for the case where $(N_\mathbf{C},N_\mathbf{D})$ = $(N_X,N_Y) = (2,1)$, as considered in this example. In fact, we can establish a general equivalence.

\paragraph{Schmidt equivalence of particle and mode states.} 

Take any single-mode basis state $\ket{n,N-n}_\mathbf{A}$, and fix a bipartition into $(N_X,N_Y)$ particles. Consider the same state after it has been split by any transformation $\hat{a}_k^\dagger\rightarrow r\hat{c}_k^\dagger + t\hat{d}_k^\dagger$, with $k=0,1,$ and $|t|^2+|r|^2=1$, followed by projection onto local particle numbers $(N_\mathbf{C},N_\mathbf{D})$. If $(N_\mathbf{C},N_\mathbf{D})=(N_X,N_Y)$ or $(N_Y,N_X)$, then the Schmidt form of the input state (in the given particle bipartition) is equivalent to the Schmidt form of the output state (in the mode bipartition).

\paragraph{Proof:}

The Schmidt form of the final state, which we denote by $\ket{n,N-n}_{(N_\mathbf{C},N_\mathbf{D})}$, can be straightforwardly but laboriously obtained by writing the input state as $\ket{n,N-n}_\mathbf{A} =\frac{\hat{a_0}^{\dagger n}\hat{a_1}^{\dagger (N-n)}}{\sqrt{n!(N-n)!}}\ket{0,0}_\mathbf{A}$, transforming $\hat{a}_k^\dagger\rightarrow r\hat{c}_k^\dagger + t\hat{d}_k^\dagger$, then projecting onto terms with fixed local particle numbers $(N_\mathbf{C},N_\mathbf{D})$, i.e., those with prefactor $\sim r^{N_\mathbf{C}}t^{N_\mathbf{D}}$; see Appendix. Once normalized, this automatically yields the Schmidt form: $\ket{n,N-n}_{(N_\mathbf{C},N_\mathbf{D})}=\sum\lambda_{n_C,n_D}\ket{u_{n_C}}_\mathbf{C}\ket{u_{n_D}}_\mathbf{D}$, where the local states of mode $\mathbf{K}=\mathbf{C},\mathbf{D}$ are second quantization basis states: $\ket{u_{n_\mathbf{K}}}=\ket{n_\mathbf{K},N_\mathbf{K}-n_\mathbf{K}}_\mathbf{K}$, and the sum is over all valid $(n_\mathbf{C},n_\mathbf{D})$ such that 
$n_\mathbf{C}+n_\mathbf{D}=n$. The Schmidt coefficients are calculated to be $\lambda_{n_\mathbf{C},n_\mathbf{D}} =\sqrt{\binom{N_\mathbf{C}}{n_\mathbf{C}}\binom{N_\mathbf{D}}{n_\mathbf{D}}/\binom{N}{n}}$.

In first-quantization, we begin with Eq. (\ref{eq:symmstate}). We subdivide this state into parts $X,Y$, where $X$ contains the pseudo-labels $1,\dots, N_X$ and $Y$ the rest (in fact, by symmetrization, the specific order will not matter). After symmetrizing, we collect terms that have the same number $n_X$ of spin-down states within $X$; $Y$ will contain the remaining $n_Y=n-n_X$. For every pair $(n_X,n_Y)$, both parts have a symmetrized form:
\begin{equation}\label{eq:bipartitesymmetrized}
	\ket{n,N-n}_\mathbf{A} = \frac{1}{\sqrt{\binom{N}{n}}}\sum_{n_X+n_Y=n} 
	\left[\mathcal{S}\ket{v_{n_X}}\right]
	\left[\mathcal{S}\ket{v_{n_Y}}\right],
\end{equation}
where $\ket{v_{n_X}}=\ket{0}_1\dots\ket{0}_{n_X}\ket{1}_{n_X+1}\dots\ket{1}_{N_X}$ and analogously for $Y$. Comparing to Eq. (1), we see that $\mathcal{S}\ket{v_{n_Z}}=\sqrt{\binom{N_Z}{n_Z}}\ket{n_Z,N_Z-n_Z}_Z$ for $Z=X,Y$. Since the states $\ket{n_{Z},N_{Z}-n_{Z}}_{Z}$ are orthonormal, this is the Schmidt form, with coefficients $\lambda_{n_X,n_Y} = \sqrt{\binom{N_X}{n_X}\binom{N_Y}{n_Y}/\binom{N}{n}}$. Thus, if $(N_\mathbf{C},N_\mathbf{D})=(N_X,N_Y)$ or $(N_Y,N_X)$, then the particle Schmidt form and the mode Schmidt form are in one-to-one correspondence. \hfill$\square$

This equivalence has strong consequences. The single-mode state $\ket{n,N-n}_\mathbf{A}$ and its two-mode equivalents $\ket{n,N-n}_{(N_\mathbf{C},N_\mathbf{D})}$ not only have the exact same entanglement structure, but the former states can be easily mapped to the latter. This holds as well for arbitrary superpositions $\ket{\phi}_\mathbf{A}=\sum_n\phi_n\ket{n,N-n}_\mathbf{A}$, since the entanglement properties within any bipartition are completely determined by the coefficients $\{\phi_n\}$ and the Schmidt structure of the basis vectors. By linearity, the algebraic correspondence also holds for mixed states. Thus, by enacting the isomorphism $\ket{n,N-n}_\mathbf{A}\mapsto\ket{n,N-n}_{(N_\mathbf{C},N_\mathbf{D})}~\forall~n,$ (with $(N_\mathbf{C},N_\mathbf{D}) = (N_X,N_Y)$), we can map any single-mode state into its two-mode version, where the entanglement structure is not only preserved, but is readily accessible. 
To emphasize, although we cannot individually access the identical particles, their overall \emph{state} is, in fact, accessible, since it can be mapped faithfully onto distinguishable mode subsystems. We will call any protocol that achieves the isomorphism 
$\ket{n,N-n}_\mathbf{A}\mapsto\ket{n,N-n}_{(N_\mathbf{C},N_\mathbf{D})}~\forall~n,$ \emph{ideal mode splitting}. 

Mode splitting does not fit in the framework of LOCC, and the process appears to `create' entanglement (this is a well-known property of beamsplitters). By the above isomorphism, the structure and amount of mode entanglement, for fixed local particle numbers, is completely determined from the input state. Thus, mode splitting is a mechanism for faithfully \emph{transferring} correlations from inaccessible identical particles onto accessible modes. If the splitting is sufficiently passive (we give  formal conditions below), all mode entanglement comes from the initial state, and no more entanglement can be generated than what appears from the $N$-particle decomposition. Finally, it is not a practical problem if a non-ideal mode splitting creates extra entanglement (e.g., by having a non-vacuum state in input mode $\mathbf{B}$); such entanglement is nevertheless a useful resource. However, we cannot interpret such entanglement as coming solely from the input state.

\paragraph{Probabilistic mode splitting and mode mixing.}

Along with tunnelling/beam-splitting, what other operations achieve the desired isomorphism? Besides the basic ability to coherently map one mode into two, there are three other important components. First, to put particle and mode entanglement on the same footing (in terms of subsystem size), and to exclude superpositions of different local particle numbers from consideration, we must project onto fixed particle numbers for each output mode. Second, the operation should not introduce any extra particles. Finally, the process should not (de-)excite the particles, so that the total number of excitations will be preserved. The reasoning for the latter two requirements is similar: such operations could lead to entanglement in the output mode when none is apparent in the input partitioning. Consider the initial state $\ket{N,0}_\mathbf{A}$, which has no pseudo-label entanglement. If the output system contains a spin-up particle (either externally added or internally excited), then output states of the form 
$~\ket{M,1}_{(N_\mathbf{C},N_\mathbf{D})}$ could appear, which are mode entangled for all $N_\mathbf{C},N_\mathbf{D}\neq0$. Obviously this entanglement is not representative of the initial state's entanglement structure. 

These basic requirements are necessary and sufficient to give the desired isomorphism --- at least probabilistically --- ensuring that the splitting itself does not contribute any entanglement \footnote{Note that one could also allow collective local unitaries (e.g., phase shifters), which would change the basis states, but not the entanglement structure in either picture.}. To show this, we consider the slightly more general situation of \emph{mode mixing}, from the input space $\mathcal{H}_\mathrm{in}=\mathrm{span}\{\ket{n,N_\mathbf{AB}-n}_{(N_\mathbf{A},N_\mathbf{B})}\}$ to the output space $\mathcal{H}_\mathrm{out}=\mathrm{span}\{\ket{n,N_\mathbf{CD}-n}_{(N_\mathbf{C},N_\mathbf{D})}\}$, where the basis states encompass all combinations with fixed global particle numbers $N_\mathbf{A}+N_\mathbf{B}=N_\mathbf{AB}$, $N_\mathbf{C}+N_\mathbf{D}=N_\mathbf{CD}.$ This extra generality will be useful for our deterministic mode-splitting protocol later. For particle excitation, we use the operators 
$\hat{J}^+_\mathbf{AB}:=\hat{a}_0\hat{a}_1^\dagger + \hat{b}_0\hat{b}_1^\dagger$ and $\hat{J}^+_\mathbf{CD}:=\hat{c}_0\hat{c}_1^\dagger + \hat{d}_0\hat{d}_1^\dagger$. The conjugate operators $\hat{J}_\mathbf{KL}^-:=(\hat{J}_\mathbf{KL}^{+})^\dagger$ model de-excitation. We denote the parameters $(N_\mathbf{A},N_\mathbf{B};N_\mathbf{C},N_\mathbf{D})$ by the shorthand $\{N\}$.

\paragraph{Theorem 1: Necessary and sufficient conditions for mode-mixing.} Let $M:\mathcal{H}_\mathrm{in}\rightarrow\mathcal{H}_\mathrm{out}$ be a linear particle-preserving map, with $N_\mathbf{AB}=N_\mathbf{CD}=N$. Then M has the effect $M\ket{n,N-n}_{(N_\mathbf{A},N_\mathbf{B})}=\sum_{N_\mathbf{C}+N_\mathbf{D}=N}C\{N\}\ket{n,N-n}_{(N_\mathbf{C},N_\mathbf{D})}~\forall~n$, where each $C\{N\}\in\mathbb{C}$, if and only if $M$ commutes with particle excitation and de-excitation, in the sense that $M\hat{J}^\pm_\mathbf{AB}=\hat{J}^\pm_\mathbf{CD}M$.

\paragraph{Proof:} 
If $M$ commutes with $\hat{J}^+$, then, by direct calculation, the matrix elements $M_{mn}^{\{N\}}:=\bra{m,N-m}_{(N_\mathbf{C},N_\mathbf{D})}M\ket{n,N-n}_{(N_\mathbf{A},N_\mathbf{B})}$ must satisfy the recurrence relation
\begin{equation}\label{eq:recurrence}
	M_{m+1,n+1}^{\{N\}} = M_{m,n}^{\{N\}}\sqrt{\frac{(n+1)(N-n)}{(m+1)(N-m)}}.
\end{equation}
As well, $M$ will commute with the operator $\hat{J}^-\hat{J}^+$, which leads to $M^{\{N\}}_{mn}\left[ n(N-n+1)-m(N-m+1) \right]= 0$. The bracketed term has roots $m=n$ and $m=N-n+1$. Combining this with the recurrence relation, we conclude that each submatrix $M^{\{N\}}$ is a multiple of the identity:
\begin{equation}\label{eq:MNform}
	M_{mn}^{\{N\}} = C\{N\}\delta_{mn}.
\end{equation}

The overall map has the form $M = \sum_{\{N\}}M^{\{N\}}$, where each $M^{\{N\}}$ satisfies Eq. (\ref{eq:MNform}), and the sum is over all ways to split up the $N$ particles in both the input and output modes. Thus, $M$ gives the desired isomorphism with some probability amplitude $C\{N\}$. Conversely, it is easy to check that any $M$ with this final form commutes with the (de-)excitation operator.
\hfill$\square$

The coefficients $C\{N\}$ play the same role as the beamsplitter-parameter combinations ($r^3$, $\sqrt{3}r^2t$, etc.) in Eq. (\ref{eq:bsoutputstate}); with suitable normalization, $|C\{N\}|^2$ is the probability that states with initial particle numbers $(N_\mathbf{A},N_\mathbf{B})$ are mapped to states with output numbers $(N_\mathbf{C},N_\mathbf{D})$. Crucially, these coefficients depend on the local particle counts, not on the states themselves. Following such maps with local particle measurements $\hat{N}_\mathbf{C}\otimes\hat{N}_\mathbf{D}$, we can probabilistically realize the desired isomorphism for any subsystem sizes. As foreshadowed by the example, \emph{any} unitary operation $\hat{a}_k^\dagger\rightarrow r\hat{c}_k^\dagger + t\hat{d}_k^\dagger$, $\hat{b}_k^\dagger\rightarrow -t^*\hat{c}_k^\dagger + r^*\hat{d}_k^\dagger$, $k=0,1$, leads to probabilistic mode mixing. In the multimode situation, this 
generalizes to polarization-independent passive optical networks. Theorem 1 can also be formulated on the $N$-particle state space; see Appendix.

\paragraph{Extraction protocol.}

Here we outline how one can extract identical-particle entanglement with unit probability using an asymptotic protocol. Take an arbitrary $N-$particle state in mode $\mathbf{A}$, $\ket{\phi_\mathrm{in}}_\mathbf{A} = \sum_{n=0}^N \phi_n \ket{n,N-n}_\mathbf{A}$, and a desired bipartition size $(N_X,N_Y)$. The protocol goes as follows: i) Apply any non-trivial mode-mixing operation $M$; ii) Measure local particle numbers $\hat{N}_\mathbf{C}\otimes \hat{N}_\mathbf{D}$. If $(N_\mathbf{C},N_\mathbf{D})=(N_X,N_Y)$ or $(N_Y,N_X)$, then the output state is $\ket{\phi_\mathrm{out}}_\mathbf{CD} = \sum_{n=0}^N \phi_n \ket{n,N-n}_{(N_X,N_Y)}$. Otherwise, repeat  step i with updated particle numbers $(N_\mathbf{A}',N_\mathbf{B}')=(N_\mathbf{C},N_\mathbf{D})$. For this protocol to work, we must show that at any iteration, with local particle counts $(N_\mathbf{A},N_\mathbf{B})$, we have a non-
zero probability to reach the desired goal $(N_X,N_Y)$ within a bounded number of steps. In fact, any non-trivial beamsplitter has $C_1=C(N_\mathbf{A},N_\mathbf{B};N,0)=\sqrt{\binom{N}{N_\mathbf{A}}}r^{N_\mathbf{A}}(-t^*)^{N_\mathbf{B}}\neq 0$ and $C_2=C(N,0;N_X,N_Y)=\sqrt{\binom{N}{N_X}}r^{N_X}t^{N_Y}\neq 0$. Thus, with only two iterations, we can guarantee an overall probability $|C_1 C_2|^2\neq 0$ of achieving the desired isomorphism. Asymptotically, we can faithfully extract any desired entangled state. 
We emphasize that during the extraction protocol, intermediate states could have quantitatively more entanglement than the final state. One should not interpret this as meaning that the protocol does not extract all available entanglement. Rather, we remember that the entanglement content is relative to the choice of bipartition of the intial state; for every choice, this protocol faithfully extracts the corresponding entangled state.

\paragraph{Relation to Spin-squeezing.}
These results reveal new operational meaning for single-mode spin-squeezed states. Squeezing information can be used to bound the expected mode entanglement, even without performing the splitting. Using only a few simple collective spin measurements \cite{korbicz05a}, we can obtain the two-particle reduced state $\rho_{pq}$, which is the same for every pseudo-label pair. We can then bound the output state's entanglement using any monogamous measure $\mathcal{E}$:
\begin{equation}
	\mathcal{E}(\rho_{\mathbf{C}:\mathbf{D}}) 
	\geq \mathcal{E}(\rho_{C_1:\mathbf{D}}) 
	\geq \sum_{j}\mathcal{E}(\rho_{C_1:D_j}) 
	= N_\mathbf{D} \mathcal{E}(\rho_{C_1:D_1}), 
\end{equation}
where the first relation follows from tracing out all qubits in $\mathbf{C}$ but $C_1$, the second represents monogamy, and the third is from symmetry; a similar inequality holds for $N_\mathbf{D}\leftrightarrow N_\mathbf{C}$. For a broad class of spin-squeezed states created by standard methods, there is a quantitative relationship between the spin-squeezing parameter $\xi^2<1$ and the concurrence \cite{wootters98a} for any $\rho_{pq}$ \cite{wang03a, wang04a, ulamorgikh01a, yin11a}. We can leverage this to bound the tangle \cite{coffman00a, osborne06a} (generalized concurrence), a measure which quantifies the usefulness of a state for bipartite channel discrimination \cite{boixo08a}:
$	\tau(\rho_{\mathbf{C}:\mathbf{D}}) \geq   \max\{N_\mathbf{C},N_\mathbf{D}\}\left[\frac{1-\xi^2}{N-1}\right]^2. $
Thus, spin-squeezed states, and the squeezing parameter $\xi^2$, acquire new operational meaning thanks to our results.

\paragraph{Conclusion.}

We have shown that identical-particle entanglement can be easily and faithfully extracted and used as a resource for standard quantum information tasks. Practically, such entanglement is naturally occuring and quite robust \cite{simon02a,stockton03a,benatti12a}. In optics, the idea to use non-classical states and beamsplitters to create entanglement has appeared many times. However, because the second quantization formalism is dominant, and because the particle superselection rules are not relevant for photons, the connection between entanglement in a discrete identical particle basis and beam-splitter generated entanglement was not previously uncovered. For massive particles, it is perhaps more natural to begin with the $N$-particle state space, but the notion of splitting and mixing modes is not as prevalent. Our results illuminate new connections between entanglement, squeezing, and indistinguishability in both scenarios. 

\begin{acknowledgments} This work was supported by the Alexander von Humboldt Foundation, the EU Integrating Project SIQS and the EU STREP EQUAM. We acknowledge G\'eza T\'oth for helpful discussions.
\end{acknowledgments}

\clearpage

\appendix
\section{Appendix}

\subsection{Appendix 1: Algebraic form of multimode and multiparticle states}

In this section, we carry out the straightforward but algebraically laborious calculation of the (generalized) Schmidt form of the state $\ket{n,N-n}_\mathbf{A}$ after a (multi-)mode splitting transformation. For simplicity, we change the notation slightly from the main text: particles with internal state $i$ in mode $\mathbf{K}$ are associated with the creation operator $\hat{a}_{i\mathbf{K}}^{\dagger}$. We also use here the same labels $\mathbf{A},\mathbf{B},\mathbf{C},\dots$ for both input and output modes, and fix a total number of modes. Symbolically, we represent the last mode in the list by  $\mathbf{Z}$, but this does not imply any specific number of modes. The initial state is thus given by
\begin{equation}\label{eq:appinputstate}
	\ket{n,N-n}_\mathbf{A} = \frac{\hat{a}_{0\mathbf{A}}^{\dagger n}\hat{a}_{1\mathbf{A}}^{\dagger (N-n)}}{\sqrt{n!(N-n)!}}\ket{\mathrm{vac}}.
\end{equation}
An arbitrary linear transformation amongst the creation operators has the form $\hat{a}_{i\mathbf{K}}^\dagger\rightarrow \sum_{j\mathbf{L}}\alpha_{i\mathbf{K},j\mathbf{L}}\hat{a}_{j\mathbf{L}}^\dagger + \beta_{i\mathbf{K},j\mathbf{L}}\hat{a}_{j\mathbf{L}}$, with $\alpha_{i\mathbf{K},j\mathbf{L}},~\beta_{i\mathbf{K},j\mathbf{L}}\in \mathbb{C}$ for $i,j\in\{0,1\}$ and $\mathbf{K},\mathbf{L}\in\{\mathbf{A},\mathbf{B},\dots \mathbf{Z}\}$. When the total number of particles is preserved and the transformation is independent of the internal state, we have the simplifications $\beta_{i\mathbf{K},j\mathbf{L}}=0$ and $\alpha_{i\mathbf{K},j\mathbf{L}}=\alpha_{\mathbf{KL}}\delta_{ij}$. Applying such a transformation to the state (\ref{eq:appinputstate}) gives
\begin{align}
	\ket{\phi_\mathrm{out}} = \frac{\left(\displaystyle\sum_{\mathbf{K}}\alpha_{\mathbf{AK}}\hat{a}_{0\mathbf{K}}^\dagger\right)^n\left(\displaystyle\sum_{\mathbf{L}}\alpha_{\mathbf{AL}}\hat{a}_{1\mathbf{L}}^\dagger\right)^{N-n}}{\sqrt{n!(N-n)!}}\ket{\mathrm{vac}},
\end{align}
where the sums are over all output modes $\mathbf{K}/\mathbf{L} = \mathbf{A},\mathbf{B},\mathbf{C},\dots, \mathbf{Z}.$ From this expression, we carry out multinomial expansions
\begin{align}
	& \left(\displaystyle\sum_{\mathbf{K}}\alpha_{\mathbf{AK}}\hat{a}_{0\mathbf{K}}^\dagger\right)^n =  \nonumber\\
	& \sum_{n_\mathbf{A}+\dots+n_\mathbf{Z}=n}\frac{n!}{n_\mathbf{A}!\cdots n_\mathbf{Z}!}\prod_{\mathbf{K}}(\alpha_{\mathbf{AK}}\hat{a}_{0\mathbf{K}}^\dagger)^{n_\mathbf{K}},\\
	& \left(\displaystyle\sum_{\mathbf{L}}\alpha_{\mathbf{AL}}\hat{a}_{1\mathbf{L}}^\dagger\right)^{N-n} = \nonumber \\
	& \sum_{m_\mathbf{A}+\dots+m_\mathbf{Z}=N-n}\frac{(N-n)!}{m_\mathbf{A}!\cdots m_\mathbf{Z}!}\prod_{\mathbf{L}}(\alpha_{\mathbf{AL}}\hat{a}_{1\mathbf{L}}^\dagger)^{m_\mathbf{L}}.
\end{align}

Since $(\hat{a}_{0\mathbf{K}}^\dagger)^{n_\mathbf{K}}(\hat{a}_{1\mathbf{K}}^\dagger)^{m_\mathbf{K}}\ket{\mathrm{vac}} = \sqrt{n_\mathbf{K}!m_\mathbf{K}!}\ket{n_\mathbf{K},m_\mathbf{K}}_\mathbf{K}$ for mode $\mathbf{K}$, the output state $\ket{\phi_\mathrm{out}}$ becomes
\begin{align}
	\sum_{\substack{n_\mathbf{A}+\dots+n_\mathbf{Z}=n\\m_\mathbf{A}+\dots+m_\mathbf{Z}=N-n}}\sqrt{\frac{n!(N-n)!}{n_\mathbf{A}!\cdots n_\mathbf{Z}! m_\mathbf{A}!\cdots m_\mathbf{Z}!}} \nonumber\\
	\bigotimes_{\mathbf{K}}\alpha_{\mathbf{AK}}^{n_\mathbf{K}+m_\mathbf{K}}\ket{n_\mathbf{K},m_\mathbf{K}}_\mathbf{K}.
\end{align}

We now group terms where each mode $\mathbf{K}$ has a fixed number of particles $N_\mathbf{K}$, i.e., $\ket{\phi_\mathrm{out}}=\sum_{N_\mathbf{A},\dots, N_\mathbf{Z}}w_{N_\mathbf{A},\dots,N_\mathbf{Z}}\ket{\phi_\mathrm{out}}_{N_\mathbf{A},\dots,N_\mathbf{Z}}$. These terms can be identified by the condition $n_\mathbf{K}+m_\mathbf{K}=N_\mathbf{K}$. We simplify the coefficients by multiplying with the unit term $\sqrt{\frac{N!N_\mathbf{A}!\cdots N_\mathbf{Z}!}{N!N_\mathbf{A}!\cdots N_\mathbf{Z}!}}$, which yields the normalized states
\begin{align}\label{eq:multimodeschmidt}
	& \ket{\phi_\mathrm{out}}_{N_\mathbf{A},\cdots,N_\mathbf{Z}} = \nonumber\\
	& \sum_{n_\mathbf{A}+\cdots+n_\mathbf{Z}=n}
	\sqrt{\frac{\binom{N_\mathbf{A}}{n_\mathbf{A}}\dots\binom{N_\mathbf{Z}}{n_\mathbf{Z}}}{\binom{N}{n}}}
	\bigotimes_{\mathbf{K}}\ket{n_\mathbf{K},N_\mathbf{K}-n_\mathbf{K}}_\mathbf{K}
\end{align}
with weights $w_{N_\mathbf{A},\dots,N_\mathbf{Z}}=\sqrt{\frac{N!}{N_\mathbf{A}!\cdots N_\mathbf{Z}!}}\prod_\mathbf{K}\alpha_{\mathbf{AK}}^{N_\mathbf{K}}$. Since the states \{$\bigotimes_\mathbf{K}\ket{n_\mathbf{K},N_\mathbf{K}-n_\mathbf{K}}_\mathbf{K}\}$ are orthonormal, Eq. (\ref{eq:multimodeschmidt}) has the form of a generalized Schmidt decomposition. For two output modes, we recover the standard Schmidt decomposition, which appeared in the main text. 

To get the multipartition form in the $N$-particle basis, we use the trick of continually splitting a single-partition into two parts. Suppose we want to partition the $N$ particles in the initial state $\ket{n,N-n}_\mathbf{A}$ into groups containing $N_{\alpha},N_\beta,N_\gamma,\dots,N_\zeta$. Without loss of generality, we form a bipartition of the first $N_\alpha$ particles and the remaining 
$N_{\overline{\alpha}} = N-N_\alpha$. By the bipartite decomposition, Eq. (4), we have
\begin{equation}\label{eq:bipartitesymmetrizedjr}
	\ket{n,N-n}_\mathbf{A} = \frac{1}{\sqrt{\binom{N}{n}}}\sum_{n_\alpha+n_{\overline{\alpha}}=n} 
	\left[\mathcal{S}\ket{v_{n_\alpha}}\right]
	\left[\mathcal{S}\ket{v_{n_{\overline{\alpha}}}}\right],
\end{equation}
where $\ket{v_{n_\alpha}}=\ket{0}_1\dots\ket{0}_{n_\alpha}\ket{1}_{n_\alpha+1}\dots\ket{1}_{N_\alpha}$, and $\ket{v_{n_{\overline{\alpha}}}}$ has an analogous form. Rewriting the state $\mathcal{S}\ket{v_{n_\alpha}}$ in second quantized form, it becomes $\sqrt{\binom{N_\alpha}{n_\alpha}}\ket{n_\alpha,N_\alpha-n_\alpha}_\alpha$.

The second partition can now be further subdivided into two parts, containing $N_\beta$ and $N_{\overline{\beta}} = N-N_\alpha-N_\beta$ particles. Continuing in this way, and rewriting each partition in second quantization, we end up with 
\begin{align}
	& \ket{\phi_\mathrm{in}}_{\mathbf{A}} = \nonumber\\
	& \sum_{n_\alpha+\cdots+n_\zeta=n}
	\sqrt{\frac{\binom{N_\alpha}{n_\alpha}\dots\binom{N_\zeta}{n_\zeta}}{\binom{N}{n}}}
	\bigotimes_{\mathbf{\kappa}}\ket{n_\mathbf{\kappa},N_\mathbf{\kappa}-n_\mathbf{\kappa}}_\mathbf{\kappa},
\end{align}
which is in one-to-one corresondence to the multimode split state in Eq. (\ref{eq:multimodeschmidt}) when $(N_\mathbf{A},\dots,N_\mathbf{Z})$ is some permutation of $(N_\alpha,\dots,N_\zeta)$. Thus, an arbitrary multimode transformation of the form $\hat{a}_{i\mathbf{K}}^\dagger\rightarrow \sum_{\mathbf{L}}\alpha_{\mathbf{K},\mathbf{L}}\hat{a}_{i\mathbf{L}}^\dagger$, $k=0,1$, will lead to states which are algebraically equivalent to their single-mode multi-particle counterparts. These transformations, familiar from optics, are passive linear optical networks with no polarization dependence.

\subsection{Appendix 2: Mode-mixing conditions in particle basis}

In this section, we present an alternate version of Theorem 1, where the conditions are instead defined relative to the $N$-particle state space. This framework might be more familiar in some cold atoms settings than the creation/annihilation operator picture. In either case, the conditions are intuitively the same. To begin, the $N$-particle basis states on the input and output spaces will be denoted by
\begin{equation}
	\ket{i;\mathbf{K}}:=\ket{i_1;\mathbf{K}_1}_1\ket{i_2;\mathbf{K}_2}_2\dots\ket{i_N;\mathbf{K}_N}_N
\end{equation}
where $i_p\in\{0,1\}$ is the $p$th entry of $i$, labeling the internal state of particle $p$. Similarly, $\mathbf{K}_p$ is the $p$th entry of $\mathbf{K}$, labeling the external mode which particle $p$ occupies. For pre-mixing states, $\mathbf{K}_p\in\{\mathbf{A},\mathbf{B}\}$; for post-mixing states, $\mathbf{K}_p\in\{\mathbf{C},\mathbf{D}\}$.

Intuitively, we begin with the same operational requirements as before, namely that the operation preserves particle numbers and does not excite the system. Particle preservation is captured simply by requiring that the operator $M$ maps between the $N$-particle state spaces $\mathcal{H}_\mathrm{in} = \mathrm{span}\{\ket{i;\mathbf{K}}~|~\mathbf{K}_p\in\{\mathbf{A},\mathbf{B}\}~\forall~p\}$ and $\mathcal{H}_\mathrm{out} = \mathrm{span}\{\ket{i;\mathbf{K}}~|~\mathbf{K}_p\in\{\mathbf{C},\mathbf{D}\}~\forall~p\}$. For (de-)excitation, we define the local operators $\hat{\sigma}^{\pm}_p$ on particle $p$ by
\begin{align}
	\hat{\sigma}_p^+\ket{i;\mathbf{K}}= & \delta_{i_p,0}\ket{i^{p+};\mathbf{K}},\\
	\hat{\sigma}_p^-\ket{i;\mathbf{K}}= & \delta_{i_p,1}\ket{i^{p-};\mathbf{K}},
\end{align}
where the $p$th element of $i^{p\pm}$ is $i_p\pm1$ and all others are the same as in $i$. If we demand that $M$ commutes with both $\hat{\sigma}_p^+$ and $\hat{\sigma}_p^-$ for every $p$, we can straightforwardly derive the following set of conditions, which hold for all $i,j,\mathbf{K},\mathbf{L}$:
\begin{align}
	\bra{i;\mathbf{K}}M\ket{j;\mathbf{L}} = 
	\begin{cases}
		\bra{i^{p+};\mathbf{K}}M\ket{j^{p+};\mathbf{L}} & \mathrm{if}~i_p = 0,~j_p = 0\\ 
		0 & \mathrm{if}~i_p = 0,~j_p = 1\\
		0 & \mathrm{if}~i_p = 1,~j_p = 0
	\end{cases}.	
\end{align}

These are actually quite stringent conditions. They tell us that M must have a local block-diagonal structure with respect to the $N$-particle state space: 
\begin{align}\label{eq:blkdiagform}
	M = & \bigotimes_{p=1}^N M^{(p)},\\
	M^{(p)} = & \sum_{i=0,1}\sum_{\mathbf{K}=\mathbf{C},\mathbf{D}}\sum_{\mathbf{L}=\mathbf{A},\mathbf{B}}C_{\mathbf{KL}}^{(p)}\ketbra{i;\mathbf{K}}{i;\mathbf{L}}.
\end{align}
Thus, $M$ simply maps particle $p$ from mode $\mathbf{L}=\mathbf{A},\mathbf{B}$ to mode $\mathbf{K}=\mathbf{C},\mathbf{D}$ with amplitude $C_{\mathbf{KL}}^{(p)}$, without considering nor changing its internal state. Even more, if the particles are truly identical, there can be no dependence on the pseudo-label $p$, so in fact, we must have $C_{\mathbf{KL}}^{(p)} = C_{\mathbf{KL}}$ and $M = [M^{(1)}]^{\otimes N}$.

\paragraph{Theorem 1a.} Let $M:\mathcal{H}_\mathrm{in}\rightarrow\mathcal{H}_\mathrm{out}$ and let $N_\mathbf{A}+N_\mathbf{B}=N_\mathbf{C}+N_\mathbf{D}=N$. If $\left[M,\hat{\sigma}_p^+\right]=\left[M,\hat{\sigma}_p^-\right]=0~\forall~p$, then $M$ has the effect $M\ket{n,N-n}_{(N_\mathbf{A},N_\mathbf{B})}=\sum_{N_\mathbf{C}+N_\mathbf{D}=N}C\{N\}\ket{n,N-n}_{(N_\mathbf{C},N_\mathbf{D})}~\forall~n$, where each $C\{N\}\in\mathbb{C}$.

\paragraph{Proof:}
First fix the number combination $\{N\}$ and consider the collective excitation operator $\hat{J}^+:= \sum_{p=1}^N \hat{\sigma}^+_p$. On two-mode symmetric states, for any local particle numbers $(N_X,N_Y)$, $\hat{J}^+$ has the following effect:
\begin{align}
	\hat{J}^+ & \ket{n,N_{XY}-n}_{(N_X,N_Y)} = \nonumber\\
	& \sqrt{n(N_{XY}-n+1)}\ket{n-1,N_{XY}-n+1}_{(N_X,N_Y)}.
\end{align}
Since $M$ commutes with each $\hat{\sigma}_p^+$, it will also commute with $\hat{J}^+$. Similar to Theorem 1, we can directly work out a recurrence relation on the matrix elements $M_{mn}^{\{N\}}$ (cf. Eq. (5)):
\begin{equation}\label{eq:recurrencejr}
	M_{m+1,n+1}^{\{N\}} = M_{m,n}^{\{N\}}\sqrt{\frac{(n+1)(N-n)}{(m+1)(N-m)}}.
\end{equation}
Furthermore, $M$ will also commute with the operator $\hat{n}_0 = \sum_{p=1}^N \hat{\sigma}_p^- \hat{\sigma}^+_p$, which has the effect
\begin{equation}
	\hat{n}_0\ket{n,N_{XY}-n}_{(N_X,N_Y)} = n\ket{n,N_{XY}-n}_{(N_X,N_Y)}.
\end{equation}
This implies that 
\begin{equation}
	M^{\{N\}}_{mn}(m-n) = 0,
\end{equation}
hence, $M^{\{N\}}$ is diagonal. Considering the recurrence relation (\ref{eq:recurrencejr}), we conclude that $M^{\{N\}}$ must be a multiple of the identity:
\begin{equation}
	M_{mn}^{\{N\}} = C\{N\}\delta_{mn}.
\end{equation}
For non-fixed bipartition sizes, we will instead have the general form $M= \sum_{\{N\}}M^{\{N\}}$ with $M^{\{N\}}$ as above. Thus, $M$ carries out the desired transformation. 
\hfill$\square$

\bibliography{mode-split-refs}

\begin{thebibliography}{10}%
\makeatletter
\providecommand \@ifxundefined [1]{%
 \ifx #1\undefined \expandafter \@firstoftwo
 \else \expandafter \@secondoftwo
\fi
}%
\providecommand \@ifnum [1]{%
 \ifnum #1\expandafter \@firstoftwo
 \else \expandafter \@secondoftwo
\fi
}%
\providecommand \enquote [1]{``#1''}%
\providecommand \bibnamefont  [1]{#1}%
\providecommand \bibfnamefont [1]{#1}%
\providecommand \citenamefont [1]{#1}%
\providecommand\href[0]{\@sanitize\@href}%
\providecommand\@href[1]{\endgroup\@@startlink{#1}\endgroup\@@href}%
\providecommand\@@href[1]{#1\@@endlink}%
\providecommand \@sanitize [0]{\begingroup\catcode`\&12\catcode`\#12\relax}%
\@ifxundefined \pdfoutput {\@firstoftwo}{%
 \@ifnum{\z@=\pdfoutput}{\@firstoftwo}{\@secondoftwo}%
}{%
 \providecommand\@@startlink[1]{\leavevmode\special{html:<a href="#1">}}%
 \providecommand\@@endlink[0]{\special{html:</a>}}%
}{%
 \providecommand\@@startlink[1]{%
  \leavevmode
  \pdfstartlink
   attr{/Border[0 0 1 ]/H/I/C[0 1 1]}%
   user{/Subtype/Link/A<</Type/Action/S/URI/URI(#1)>>}%
  \relax
 }%
 \providecommand\@@endlink[0]{\pdfendlink}%
}%
\providecommand \url  [0]{\begingroup\@sanitize \@url }%
\providecommand \@url [1]{\endgroup\@href {#1}{\urlprefix}}%
\providecommand \urlprefix [0]{URL }%
\providecommand \Eprint[0]{\href }%
\@ifxundefined \urlstyle {%
  \providecommand \doi [1]{doi:\discretionary{}{}{}#1}%
}{%
  \providecommand \doi [0]{doi:\discretionary{}{}{}\begingroup
  \urlstyle{rm}\Url }%
}%
\providecommand \doibase [0]{http://dx.doi.org/}%
\providecommand \Doi[1]{\href{\doibase#1}}%
\providecommand \bibAnnote [3]{%
  \BibitemShut{#1}%
  \begin{quotation}\noindent
    \textsc{Key:}\ #2\\\textsc{Annotation:}\ #3%
  \end{quotation}%
}%
\providecommand \bibAnnoteFile [2]{%
  \IfFileExists{#2}{\bibAnnote {#1} {#2} {\input{#2}}}{}%
}%
\providecommand \typeout [0]{\immediate \write \m@ne }%
\providecommand \selectlanguage [0]{\@gobble}%
\providecommand \bibinfo [0]{\@secondoftwo}%
\providecommand \bibfield [0]{\@secondoftwo}%
\providecommand \translation [1]{[#1]}%
\providecommand \BibitemOpen[0]{}%
\providecommand \bibitemStop [0]{}%
\providecommand \bibitemNoStop [0]{.\EOS\space}%
\providecommand \EOS [0]{\spacefactor3000\relax}%
\providecommand \BibitemShut [1]{\csname bibitem#1\endcsname}%
\bibitem{feynmann10a}%
  \BibitemOpen
  \bibfield{author}{%
  \bibinfo {author} {\bibfnamefont{R.~P.}\ \bibnamefont{Feynman}}, \bibinfo
  {author} {\bibfnamefont{A.~R.}\ \bibnamefont{Hibbs}},\ and\ \bibinfo {author}
  {\bibfnamefont{D.~F.}\ \bibnamefont{Styer}},\ }%
  \emph{\bibinfo {title} {Quantum mechanics and path integrals}}\ (\bibinfo
  {publisher} {Dover},\ \bibinfo {address} {New York},\ \bibinfo {year}
  {2010})%
  \bibAnnoteFile{NoStop}{feynmann10a}%
\bibitem{plenio07a}%
  \BibitemOpen
  \bibfield{author}{%
  \bibinfo {author} {\bibfnamefont{M.~B.}\ \bibnamefont{Plenio}}\ and\ \bibinfo
  {author} {\bibfnamefont{S.}~\bibnamefont{Virmani}},\ }%
  \bibfield{journal}{%
  \bibinfo {journal} {Quantum Inf. Comput.}\ }%
  \textbf{\bibinfo {volume} {7}},\ \bibinfo {pages} {1} (\bibinfo {year}
  {2007})%
  \bibAnnoteFile{NoStop}{plenio07a}%
\bibitem{esteve08a}%
  \BibitemOpen
  \bibfield{author}{%
  \bibinfo {author} {\bibfnamefont{J.}~\bibnamefont{Esteve}}, \bibinfo {author}
  {\bibfnamefont{C.}~\bibnamefont{Gross}}, \bibinfo {author}
  {\bibfnamefont{A.}~\bibnamefont{Weller}}, \bibinfo {author}
  {\bibfnamefont{S.}~\bibnamefont{Giovanazzi}},\ and\ \bibinfo {author}
  {\bibfnamefont{M.}~\bibnamefont{Oberthaler}},\ }%
  \bibfield{journal}{%
  \Doi{10.1038/nature07332}{\bibinfo {journal} {Nature}}\ }%
  \textbf{\bibinfo {volume} {455}},\ \bibinfo {pages} {1216} (\bibinfo {year}
  {2008})%
  \bibAnnoteFile{NoStop}{esteve08a}%
\bibitem{riedel10a}%
  \BibitemOpen
  \bibfield{author}{%
  \bibinfo {author} {\bibfnamefont{M.~F.}\ \bibnamefont{Riedel}}, \bibinfo
  {author} {\bibfnamefont{P.}~\bibnamefont{B\"{o}hi}}, \bibinfo {author}
  {\bibfnamefont{Y.}~\bibnamefont{Li}}, \bibinfo {author}
  {\bibfnamefont{T.~W.}\ \bibnamefont{H\"{a}nsch}}, \bibinfo {author}
  {\bibfnamefont{A.}~\bibnamefont{Sinatra}},\ and\ \bibinfo {author}
  {\bibfnamefont{P.}~\bibnamefont{Treutlein}},\ }%
  \bibfield{journal}{%
  \Doi{10.1038/nature08988}{\bibinfo {journal} {Nature}}\ }%
  \textbf{\bibinfo {volume} {464}},\ \bibinfo {pages} {1170} (\bibinfo {year}
  {2010})%
  \bibAnnoteFile{NoStop}{riedel10a}%
\bibitem{gross10a}%
  \BibitemOpen
  \bibfield{author}{%
  \bibinfo {author} {\bibfnamefont{C.}~\bibnamefont{Gross}}, \bibinfo {author}
  {\bibfnamefont{T.}~\bibnamefont{Zibold}}, \bibinfo {author}
  {\bibfnamefont{E.}~\bibnamefont{Nicklas}}, \bibinfo {author}
  {\bibfnamefont{J.}~\bibnamefont{Estève}},\ and\ \bibinfo {author}
  {\bibfnamefont{M.~K.}\ \bibnamefont{Oberthaler}},\ }%
  \bibfield{journal}{%
  \Doi{10.1038/nature08919}{\bibinfo {journal} {Nature}}\ }%
  \textbf{\bibinfo {volume} {464}},\ \bibinfo {pages} {1165} (\bibinfo {year}
  {2010})%
  \bibAnnoteFile{NoStop}{gross10a}%
\bibitem{sorensen01a}%
  \BibitemOpen
  \bibfield{author}{%
  \bibinfo {author} {\bibfnamefont{A.}~\bibnamefont{S\o{}rensen}}, \bibinfo
  {author} {\bibfnamefont{L.}~\bibnamefont{Duan}}, \bibinfo {author}
  {\bibfnamefont{J.}~\bibnamefont{Cirac}},\ and\ \bibinfo {author}
  {\bibfnamefont{P.}~\bibnamefont{Zoller}},\ }%
  \bibfield{journal}{%
  \Doi{10.1038/35051038}{\bibinfo {journal} {Nature}}\ }%
  \textbf{\bibinfo {volume} {409}},\ \bibinfo {pages} {63} (\bibinfo {year}
  {2001})%
  \bibAnnoteFile{NoStop}{sorensen01a}%
\bibitem{sorensen01b}%
  \BibitemOpen
  \bibfield{author}{%
  \bibinfo {author} {\bibfnamefont{A.~S.}\ \bibnamefont{S\o{}rensen}}\ and\
  \bibinfo {author} {\bibfnamefont{K.}~\bibnamefont{M\o{}lmer}},\ }%
  \bibfield{journal}{%
  \Doi{10.1103/PhysRevLett.86.4431}{\bibinfo {journal} {Phys. Rev. Lett.}}\ }%
  \textbf{\bibinfo {volume} {86}},\ \bibinfo {pages} {4431} (\bibinfo {year}
  {2001})%
  \bibAnnoteFile{NoStop}{sorensen01b}%
\bibitem{omar02a}%
  \BibitemOpen
  \bibfield{author}{%
  \bibinfo {author} {\bibfnamefont{Y.}~\bibnamefont{Omar}}, \bibinfo {author}
  {\bibfnamefont{N.}~\bibnamefont{Paunkovi\ifmmode~\acute{c}\else \'{c}\fi{}}},
  \bibinfo {author} {\bibfnamefont{S.}~\bibnamefont{Bose}},\ and\ \bibinfo
  {author} {\bibfnamefont{V.}~\bibnamefont{Vedral}},\ }%
  \bibfield{journal}{%
  \Doi{10.1103/PhysRevA.65.062305}{\bibinfo {journal} {Phys. Rev. A}}\ }%
  \textbf{\bibinfo {volume} {65}},\ \bibinfo {pages} {062305} (\bibinfo {year}
  {2002})%
  \bibAnnoteFile{NoStop}{omar02a}%
\bibitem{micheli03a}%
  \BibitemOpen
  \bibfield{author}{%
  \bibinfo {author} {\bibfnamefont{A.}~\bibnamefont{Micheli}}, \bibinfo
  {author} {\bibfnamefont{D.}~\bibnamefont{Jaksch}}, \bibinfo {author}
  {\bibfnamefont{J.~I.}\ \bibnamefont{Cirac}},\ and\ \bibinfo {author}
  {\bibfnamefont{P.}~\bibnamefont{Zoller}},\ }%
  \bibfield{journal}{%
  \Doi{10.1103/PhysRevA.67.013607}{\bibinfo {journal} {Phys. Rev. A}}\ }%
  \textbf{\bibinfo {volume} {67}},\ \bibinfo {pages} {013607} (\bibinfo {year}
  {2003})%
  \bibAnnoteFile{NoStop}{micheli03a}%
\bibitem{hines03a}%
  \BibitemOpen
  \bibfield{author}{%
  \bibinfo {author} {\bibfnamefont{A.~P.}\ \bibnamefont{Hines}}, \bibinfo
  {author} {\bibfnamefont{R.~H.}\ \bibnamefont{McKenzie}},\ and\ \bibinfo
  {author} {\bibfnamefont{G.~J.}\ \bibnamefont{Milburn}},\ }%
  \bibfield{journal}{%
  \Doi{10.1103/PhysRevA.67.013609}{\bibinfo {journal} {Phys. Rev. A}}\ }%
  \textbf{\bibinfo {volume} {67}},\ \bibinfo {pages} {013609} (\bibinfo {year}
  {2003})%
  \bibAnnoteFile{NoStop}{hines03a}%
\bibitem{ghirardi04a}%
  \BibitemOpen
  \bibfield{author}{%
  \bibinfo {author} {\bibfnamefont{G.~C.}\ \bibnamefont{Ghirardi}}\ and\
  \bibinfo {author} {\bibfnamefont{L.}~\bibnamefont{Marinatto}},\ }%
  \bibfield{journal}{%
  \Doi{10.1103/PhysRevA.70.012109}{\bibinfo {journal} {Phys. Rev. A}}\ }%
  \textbf{\bibinfo {volume} {70}},\ \bibinfo {pages} {012109} (\bibinfo {year}
  {2004})%
  \bibAnnoteFile{NoStop}{ghirardi04a}%
\bibitem{korbicz05a}%
  \BibitemOpen
  \bibfield{author}{%
  \bibinfo {author} {\bibfnamefont{J.~K.}\ \bibnamefont{Korbicz}}, \bibinfo
  {author} {\bibfnamefont{J.~I.}\ \bibnamefont{Cirac}},\ and\ \bibinfo {author}
  {\bibfnamefont{M.}~\bibnamefont{Lewenstein}},\ }%
  \bibfield{journal}{%
  \Doi{10.1103/PhysRevLett.95.120502}{\bibinfo {journal} {Phys. Rev. Lett.}}\
  }%
  \textbf{\bibinfo {volume} {95}},\ \bibinfo {pages} {120502} (\bibinfo {year}
  {2005})%
  \bibAnnoteFile{NoStop}{korbicz05a}%
\bibitem{cavalcanti07a}%
  \BibitemOpen
  \bibfield{author}{%
  \bibinfo {author} {\bibfnamefont{D.}~\bibnamefont{Cavalcanti}}, \bibinfo
  {author} {\bibfnamefont{L.~M.}\ \bibnamefont{Malard}}, \bibinfo {author}
  {\bibfnamefont{F.~M.}\ \bibnamefont{Matinaga}}, \bibinfo {author}
  {\bibfnamefont{M.~O.}\ \bibnamefont{Terra~Cunha}},\ and\ \bibinfo {author}
  {\bibfnamefont{M.~F.}\ \bibnamefont{Santos}},\ }%
  \bibfield{journal}{%
  \Doi{10.1103/PhysRevB.76.113304}{\bibinfo {journal} {Phys. Rev. B}}\ }%
  \textbf{\bibinfo {volume} {76}},\ \bibinfo {pages} {113304} (\bibinfo {year}
  {2007})%
  \bibAnnoteFile{NoStop}{cavalcanti07a}%
\bibitem{hyllus10b}%
  \BibitemOpen
  \bibfield{author}{%
  \bibinfo {author} {\bibfnamefont{P.}~\bibnamefont{Hyllus}}, \bibinfo {author}
  {\bibfnamefont{L.}~\bibnamefont{Pezz\'e}},\ and\ \bibinfo {author}
  {\bibfnamefont{A.}~\bibnamefont{Smerzi}},\ }%
  \bibfield{journal}{%
  \Doi{10.1103/PhysRevLett.105.120501}{\bibinfo {journal} {Phys. Rev. Lett.}}\
  }%
  \textbf{\bibinfo {volume} {105}},\ \bibinfo {pages} {120501} (\bibinfo {year}
  {2010})%
  \bibAnnoteFile{NoStop}{hyllus10b}%
\bibitem{hyllus12a}%
  \BibitemOpen
  \bibfield{author}{%
  \bibinfo {author} {\bibfnamefont{P.}~\bibnamefont{Hyllus}}, \bibinfo {author}
  {\bibfnamefont{L.}~\bibnamefont{Pezz\'e}}, \bibinfo {author}
  {\bibfnamefont{A.}~\bibnamefont{Smerzi}},\ and\ \bibinfo {author}
  {\bibfnamefont{G.}~\bibnamefont{T\'oth}},\ }%
  \bibfield{journal}{%
  \Doi{10.1103/PhysRevA.86.012337}{\bibinfo {journal} {Phys. Rev. A}}\ }%
  \textbf{\bibinfo {volume} {86}},\ \bibinfo {pages} {012337} (\bibinfo {year}
  {2012})%
  \bibAnnoteFile{NoStop}{hyllus12a}%
\bibitem{benatti11a}%
  \BibitemOpen
  \bibfield{author}{%
  \bibinfo {author} {\bibfnamefont{F.}~\bibnamefont{Benatti}}, \bibinfo
  {author} {\bibfnamefont{R.}~\bibnamefont{Floreanini}},\ and\ \bibinfo
  {author} {\bibfnamefont{U.}~\bibnamefont{Marzolino}},\ }%
  \bibfield{journal}{%
  \Doi{doi:10.1088/0953-4075/44/9/091001}{\bibinfo {journal} {J. Phys. B: At.
  Mol. Opt. Phys.}}\ }%
  \textbf{\bibinfo {volume} {44}},\ \bibinfo {pages} {091001} (\bibinfo {year}
  {2011})%
  \bibAnnoteFile{NoStop}{benatti11a}%
\bibitem{he11a}%
  \BibitemOpen
  \bibfield{author}{%
  \bibinfo {author} {\bibfnamefont{Q.~Y.}\ \bibnamefont{He}}, \bibinfo {author}
  {\bibfnamefont{M.~D.}\ \bibnamefont{Reid}}, \bibinfo {author}
  {\bibfnamefont{T.~G.}\ \bibnamefont{Vaughan}}, \bibinfo {author}
  {\bibfnamefont{C.}~\bibnamefont{Gross}}, \bibinfo {author}
  {\bibfnamefont{M.}~\bibnamefont{Oberthaler}},\ and\ \bibinfo {author}
  {\bibfnamefont{P.~D.}\ \bibnamefont{Drummond}},\ }%
  \bibfield{journal}{%
  \Doi{10.1103/PhysRevLett.106.120405}{\bibinfo {journal} {Phys. Rev. Lett.}}\
  }%
  \textbf{\bibinfo {volume} {106}},\ \bibinfo {pages} {120405} (\bibinfo {year}
  {2011})%
  \bibAnnoteFile{NoStop}{he11a}%
\bibitem{dalton13a}%
  \BibitemOpen
  \bibfield{author}{%
  \bibinfo {author} {\bibfnamefont{B.~J.}\ \bibnamefont{Dalton}}, \bibinfo
  {author} {\bibfnamefont{L.}~\bibnamefont{Heaney}}, \bibinfo {author}
  {\bibfnamefont{J.}~\bibnamefont{Goold}}, \bibinfo {author}
  {\bibfnamefont{T.}~\bibnamefont{Busch}},\ and\ \bibinfo {author}
  {\bibfnamefont{B.~M.}\ \bibnamefont{Garraway}},\ }%
  \Eprint{http://arxiv.org/abs/1305.0788}{arXiv:1305.0788}%
  \bibAnnoteFile{NoStop}{dalton13a}%
\bibitem{dalton14a}%
  \BibitemOpen
  \bibfield{author}{%
  \bibinfo {author} {\bibfnamefont{B.~J.}\ \bibnamefont{Dalton}}, \bibinfo
  {author} {\bibfnamefont{L.}~\bibnamefont{Heaney}}, \bibinfo {author}
  {\bibfnamefont{J.}~\bibnamefont{Goold}}, \bibinfo {author}
  {\bibfnamefont{B.~M.}\ \bibnamefont{Garraway}},\ and\ \bibinfo {author}
  {\bibfnamefont{T.}~\bibnamefont{Busch}},\ }%
  \bibfield{journal}{%
  \bibinfo {journal} {New J. Phys.}\ }%
  \textbf{\bibinfo {volume} {16}},\ \bibinfo {pages} {013026} (\bibinfo {year}
  {2014})%
  \bibAnnoteFile{NoStop}{dalton14a}%
\bibitem{kitagawa93a}%
  \BibitemOpen
  \bibfield{author}{%
  \bibinfo {author} {\bibfnamefont{M.}~\bibnamefont{Kitagawa}}\ and\ \bibinfo
  {author} {\bibfnamefont{M.}~\bibnamefont{Ueda}},\ }%
  \bibfield{journal}{%
  \Doi{10.1103/PhysRevA.47.5138}{\bibinfo {journal} {Phys. Rev. A}}\ }%
  \textbf{\bibinfo {volume} {47}},\ \bibinfo {pages} {5138} (\bibinfo {year}
  {1993})%
  \bibAnnoteFile{NoStop}{kitagawa93a}%
\bibitem{wineland94a}%
  \BibitemOpen
  \bibfield{author}{%
  \bibinfo {author} {\bibfnamefont{D.~J.}\ \bibnamefont{Wineland}}, \bibinfo
  {author} {\bibfnamefont{J.~J.}\ \bibnamefont{Bollinger}}, \bibinfo {author}
  {\bibfnamefont{W.~M.}\ \bibnamefont{Itano}},\ and\ \bibinfo {author}
  {\bibfnamefont{D.~J.}\ \bibnamefont{Heinzen}},\ }%
  \bibfield{journal}{%
  \Doi{10.1103/PhysRevA.50.67}{\bibinfo {journal} {Phys. Rev. A}}\ }%
  \textbf{\bibinfo {volume} {50}},\ \bibinfo {pages} {67} (\bibinfo {year}
  {1994})%
  \bibAnnoteFile{NoStop}{wineland94a}%
\bibitem{huelga97a}%
  \BibitemOpen
  \bibfield{author}{%
  \bibinfo {author} {\bibfnamefont{S.~F.}\ \bibnamefont{Huelga}}, \bibinfo
  {author} {\bibfnamefont{C.}~\bibnamefont{Macchiavello}}, \bibinfo {author}
  {\bibfnamefont{T.}~\bibnamefont{Pellizzari}}, \bibinfo {author}
  {\bibfnamefont{A.~K.}\ \bibnamefont{Ekert}}, \bibinfo {author}
  {\bibfnamefont{M.~B.}\ \bibnamefont{Plenio}},\ and\ \bibinfo {author}
  {\bibfnamefont{J.~I.}\ \bibnamefont{Cirac}},\ }%
  \bibfield{journal}{%
  \Doi{10.1103/PhysRevLett.79.3865}{\bibinfo {journal} {Phys. Rev. Lett.}}\ }%
  \textbf{\bibinfo {volume} {79}},\ \bibinfo {pages} {3865} (\bibinfo {year}
  {1997})%
  \bibAnnoteFile{NoStop}{huelga97a}%
\bibitem{ichikawa08a}%
  \BibitemOpen
  \bibfield{author}{%
  \bibinfo {author} {\bibfnamefont{T.}~\bibnamefont{Ichikawa}}, \bibinfo
  {author} {\bibfnamefont{T.}~\bibnamefont{Sasaki}}, \bibinfo {author}
  {\bibfnamefont{I.}~\bibnamefont{Tsutsui}},\ and\ \bibinfo {author}
  {\bibfnamefont{N.}~\bibnamefont{Yonezawa}},\ }%
  \bibfield{journal}{%
  \Doi{10.1103/PhysRevA.78.052105}{\bibinfo {journal} {Phys. Rev. A}}\ }%
  \textbf{\bibinfo {volume} {78}},\ \bibinfo {pages} {052105} (\bibinfo {year}
  {2008})%
  \bibAnnoteFile{NoStop}{ichikawa08a}%
\bibitem{wei10a}%
  \BibitemOpen
  \bibfield{author}{%
  \bibinfo {author} {\bibfnamefont{T.-C.}\ \bibnamefont{Wei}},\ }%
  \bibfield{journal}{%
  \Doi{10.1103/PhysRevA.81.054102}{\bibinfo {journal} {Phys. Rev. A}}\ }%
  \textbf{\bibinfo {volume} {81}},\ \bibinfo {pages} {054102} (\bibinfo {year}
  {2010})%
  \bibAnnoteFile{NoStop}{wei10a}%
\bibitem{schliemann01a}%
  \BibitemOpen
  \bibfield{author}{%
  \bibinfo {author} {\bibfnamefont{J.}~\bibnamefont{Schliemann}}, \bibinfo
  {author} {\bibfnamefont{D.}~\bibnamefont{Loss}},\ and\ \bibinfo {author}
  {\bibfnamefont{A.~H.}\ \bibnamefont{MacDonald}},\ }%
  \bibfield{journal}{%
  \Doi{10.1103/PhysRevB.63.085311}{\bibinfo {journal} {Phys. Rev. B}}\ }%
  \textbf{\bibinfo {volume} {63}},\ \bibinfo {pages} {085311} (\bibinfo {year}
  {2001})%
  \bibAnnoteFile{NoStop}{schliemann01a}%
\bibitem{schliemann01b}%
  \BibitemOpen
  \bibfield{author}{%
  \bibinfo {author} {\bibfnamefont{J.}~\bibnamefont{Schliemann}}, \bibinfo
  {author} {\bibfnamefont{J.~I.}\ \bibnamefont{Cirac}}, \bibinfo {author}
  {\bibfnamefont{M.}~\bibnamefont{Ku\ifmmode~\acute{s}\else \'{s}\fi{}}},
  \bibinfo {author} {\bibfnamefont{M.}~\bibnamefont{Lewenstein}},\ and\
  \bibinfo {author} {\bibfnamefont{D.}~\bibnamefont{Loss}},\ }%
  \bibfield{journal}{%
  \Doi{10.1103/PhysRevA.64.022303}{\bibinfo {journal} {Phys. Rev. A}}\ }%
  \textbf{\bibinfo {volume} {64}},\ \bibinfo {pages} {022303} (\bibinfo {year}
  {2001})%
  \bibAnnoteFile{NoStop}{schliemann01b}%
\bibitem{eckert02a}%
  \BibitemOpen
  \bibfield{author}{%
  \bibinfo {author} {\bibfnamefont{K.}~\bibnamefont{Eckert}}, \bibinfo {author}
  {\bibfnamefont{J.}~\bibnamefont{Schliemann}}, \bibinfo {author}
  {\bibfnamefont{D.}~\bibnamefont{Bru{\ss}}},\ and\ \bibinfo {author}
  {\bibfnamefont{M.}~\bibnamefont{Lewenstein}},\ }%
  \bibfield{journal}{%
  \Doi{10.1006/aphy.2002.6268}{\bibinfo {journal} {Ann. Phys.}}\ }%
  \textbf{\bibinfo {volume} {299}},\ \bibinfo {pages} {88} (\bibinfo {year}
  {2002})%
  \bibAnnoteFile{NoStop}{eckert02a}%
\bibitem{ghirardi77a}%
  \BibitemOpen
  \bibfield{author}{%
  \bibinfo {author} {\bibfnamefont{G.~C.}\ \bibnamefont{Ghirardi}}, \bibinfo
  {author} {\bibfnamefont{A.}~\bibnamefont{Rimini}}, \bibinfo {author}
  {\bibfnamefont{T.}~\bibnamefont{Weber}},\ and\ \bibinfo {author}
  {\bibfnamefont{C.}~\bibnamefont{Omero}},\ }%
  \bibfield{journal}{%
  \Doi{10.1007/BF02738183}{\bibinfo {journal} {Il Nuovo Cimento B}}\ }%
  \textbf{\bibinfo {volume} {39}},\ \bibinfo {pages} {130} (\bibinfo {year}
  {1977})%
  \bibAnnoteFile{NoStop}{ghirardi77a}%
\bibitem{ghirardi02a}%
  \BibitemOpen
  \bibfield{author}{%
  \bibinfo {author} {\bibfnamefont{G.~C.}\ \bibnamefont{Ghirardi}}, \bibinfo
  {author} {\bibfnamefont{L.}~\bibnamefont{Marinatto}},\ and\ \bibinfo {author}
  {\bibfnamefont{T.}~\bibnamefont{Weber}},\ }%
  \bibfield{journal}{%
  \Doi{10.1023/A:1015439502289}{\bibinfo {journal} {J. Stat. Phys.}}\ }%
  \textbf{\bibinfo {volume} {108}},\ \bibinfo {pages} {49} (\bibinfo {year}
  {2002})%
  \bibAnnoteFile{NoStop}{ghirardi02a}%
\bibitem{paskauskas01a}%
  \BibitemOpen
  \bibfield{author}{%
  \bibinfo {author} {\bibfnamefont{R.}~\bibnamefont{Pa\v{s}kauskas}}\ and\
  \bibinfo {author} {\bibfnamefont{L.}~\bibnamefont{You}},\ }%
  \bibfield{journal}{%
  \Doi{10.1103/PhysRevA.64.042310}{\bibinfo {journal} {Phys. Rev. A}}\ }%
  \textbf{\bibinfo {volume} {64}},\ \bibinfo {pages} {042310} (\bibinfo {year}
  {2001})%
  \bibAnnoteFile{NoStop}{paskauskas01a}%
\bibitem{zanardi02a}%
  \BibitemOpen
  \bibfield{author}{%
  \bibinfo {author} {\bibfnamefont{P.}~\bibnamefont{Zanardi}},\ }%
  \bibfield{journal}{%
  \Doi{10.1103/PhysRevA.65.042101}{\bibinfo {journal} {Phys. Rev. A}}\ }%
  \textbf{\bibinfo {volume} {65}},\ \bibinfo {pages} {042101} (\bibinfo {year}
  {2002})%
  \bibAnnoteFile{NoStop}{zanardi02a}%
\bibitem{shi03a}%
  \BibitemOpen
  \bibfield{author}{%
  \bibinfo {author} {\bibfnamefont{Y.}~\bibnamefont{Shi}},\ }%
  \bibfield{journal}{%
  \Doi{10.1103/PhysRevA.67.024301}{\bibinfo {journal} {Phys. Rev. A}}\ }%
  \textbf{\bibinfo {volume} {67}},\ \bibinfo {pages} {024301} (\bibinfo {year}
  {2003})%
  \bibAnnoteFile{NoStop}{shi03a}%
\bibitem{barnum04a}%
  \BibitemOpen
  \bibfield{author}{%
  \bibinfo {author} {\bibfnamefont{H.}~\bibnamefont{Barnum}}, \bibinfo {author}
  {\bibfnamefont{E.}~\bibnamefont{Knill}}, \bibinfo {author}
  {\bibfnamefont{G.}~\bibnamefont{Ortiz}}, \bibinfo {author}
  {\bibfnamefont{R.}~\bibnamefont{Somma}},\ and\ \bibinfo {author}
  {\bibfnamefont{L.}~\bibnamefont{Viola}},\ }%
  \bibfield{journal}{%
  \Doi{10.1103/PhysRevLett.92.107902}{\bibinfo {journal} {Phys. Rev. Lett.}}\
  }%
  \textbf{\bibinfo {volume} {92}},\ \bibinfo {pages} {107902} (\bibinfo {year}
  {2004})%
  \bibAnnoteFile{NoStop}{barnum04a}%
\bibitem{barnum05a}%
  \BibitemOpen
  \bibfield{author}{%
  \bibinfo {author} {\bibfnamefont{H.}~\bibnamefont{Barnum}}, \bibinfo {author}
  {\bibfnamefont{G.}~\bibnamefont{Ortiz}}, \bibinfo {author}
  {\bibfnamefont{R.}~\bibnamefont{Somma}},\ and\ \bibinfo {author}
  {\bibfnamefont{L.}~\bibnamefont{Viola}},\ }%
  \bibfield{journal}{%
  \Doi{10.1007/s10773-005-8009-z}{\bibinfo {journal} {Int. J. Theor. Phys.}}\
  }%
  \textbf{\bibinfo {volume} {44}},\ \bibinfo {pages} {2127} (\bibinfo {year}
  {2005})%
  \bibAnnoteFile{NoStop}{barnum05a}%
\bibitem{levay05a}%
  \BibitemOpen
  \bibfield{author}{%
  \bibinfo {author} {\bibfnamefont{P.}~\bibnamefont{L\'evay}}, \bibinfo
  {author} {\bibfnamefont{S.}~\bibnamefont{Nagy}},\ and\ \bibinfo {author}
  {\bibfnamefont{J.}~\bibnamefont{Pipek}},\ }%
  \bibfield{journal}{%
  \Doi{10.1103/PhysRevA.72.022302}{\bibinfo {journal} {Phys. Rev. A}}\ }%
  \textbf{\bibinfo {volume} {72}},\ \bibinfo {pages} {022302} (\bibinfo {year}
  {2005})%
  \bibAnnoteFile{NoStop}{levay05a}%
\bibitem{plastino09a}%
  \BibitemOpen
  \bibfield{author}{%
  \bibinfo {author} {\bibfnamefont{A.}~\bibnamefont{Plastino}}, \bibinfo
  {author} {\bibfnamefont{D.}~\bibnamefont{Manzano}},\ and\ \bibinfo {author}
  {\bibfnamefont{J.}~\bibnamefont{Dehesa}},\ }%
  \bibfield{journal}{%
  \Doi{10.1209/0295-5075/86/20005}{\bibinfo {journal} {Europhys. Lett.}}\ }%
  \textbf{\bibinfo {volume} {86}},\ \bibinfo {pages} {20005} (\bibinfo {year}
  {2009})%
  \bibAnnoteFile{NoStop}{plastino09a}%
\bibitem{tichy13a}%
  \BibitemOpen
  \bibfield{author}{%
  \bibinfo {author} {\bibfnamefont{M.}~\bibnamefont{Tichy}}, \bibinfo {author}
  {\bibfnamefont{F.}~\bibnamefont{de~Melo}}, \bibinfo {author}
  {\bibfnamefont{M.}~\bibnamefont{Ku\'{s}}}, \bibinfo {author}
  {\bibfnamefont{F.}~\bibnamefont{Mintert}},\ and\ \bibinfo {author}
  {\bibfnamefont{A.}~\bibnamefont{Buchleitner}},\ }%
  \bibfield{journal}{%
  \Doi{10.1002/prop.201200079}{\bibinfo {journal} {Fortschr. Phys.}}\ }%
  \textbf{\bibinfo {volume} {61}},\ \bibinfo {pages} {225} (\bibinfo {year}
  {2013})%
  \bibAnnoteFile{NoStop}{tichy13a}%
\bibitem{zanardi04a}%
  \BibitemOpen
  \bibfield{author}{%
  \bibinfo {author} {\bibfnamefont{P.}~\bibnamefont{Zanardi}}, \bibinfo
  {author} {\bibfnamefont{D.~A.}\ \bibnamefont{Lidar}},\ and\ \bibinfo {author}
  {\bibfnamefont{S.}~\bibnamefont{Lloyd}},\ }%
  \bibfield{journal}{%
  \Doi{10.1103/PhysRevLett.92.060402}{\bibinfo {journal} {Phys. Rev. Lett.}}\
  }%
  \textbf{\bibinfo {volume} {92}},\ \bibinfo {pages} {060402} (\bibinfo {year}
  {2004})%
  \bibAnnoteFile{NoStop}{zanardi04a}%
\bibitem{sasaki11a}%
  \BibitemOpen
  \bibfield{author}{%
  \bibinfo {author} {\bibfnamefont{T.}~\bibnamefont{Sasaki}}, \bibinfo {author}
  {\bibfnamefont{T.}~\bibnamefont{Ichikawa}},\ and\ \bibinfo {author}
  {\bibfnamefont{I.}~\bibnamefont{Tsutsui}},\ }%
  \bibfield{journal}{%
  \Doi{10.1103/PhysRevA.83.012113}{\bibinfo {journal} {Phys. Rev. A}}\ }%
  \textbf{\bibinfo {volume} {83}},\ \bibinfo {pages} {012113} (\bibinfo {year}
  {2011})%
  \bibAnnoteFile{NoStop}{sasaki11a}%
\bibitem{balachandran13a}%
  \BibitemOpen
  \bibfield{author}{%
  \bibinfo {author} {\bibfnamefont{A.~P.}\ \bibnamefont{Balachandran}},
  \bibinfo {author} {\bibfnamefont{T.~R.}\ \bibnamefont{Govindarajan}},
  \bibinfo {author} {\bibfnamefont{A.~R.}\ \bibnamefont{de~Queiroz}},\ and\
  \bibinfo {author} {\bibfnamefont{A.~F.}\ \bibnamefont{Reyes-Lega}},\ }%
  \bibfield{journal}{%
  \Doi{10.1103/PhysRevLett.110.080503}{\bibinfo {journal} {Phys. Rev. Lett.}}\
  }%
  \textbf{\bibinfo {volume} {110}},\ \bibinfo {pages} {080503} (\bibinfo {year}
  {2013})%
  \bibAnnoteFile{NoStop}{balachandran13a}%
\bibitem{pezze09a}%
  \BibitemOpen
  \bibfield{author}{%
  \bibinfo {author} {\bibfnamefont{L.}~\bibnamefont{Pezz\'e}}\ and\ \bibinfo
  {author} {\bibfnamefont{A.}~\bibnamefont{Smerzi}},\ }%
  \bibfield{journal}{%
  \Doi{10.1103/PhysRevLett.102.100401}{\bibinfo {journal} {Phys. Rev. Lett.}}\
  }%
  \textbf{\bibinfo {volume} {102}},\ \bibinfo {pages} {100401} (\bibinfo {year}
  {2009})%
  \bibAnnoteFile{NoStop}{pezze09a}%
\bibitem{Note1}%
  \BibitemOpen
  \bibinfo {note} {Note that for metrology applications, it is not essential to
  have access to the individual particles.}%
  \bibAnnoteFile{Stop}{Note1}%
\bibitem{wiseman03a}%
  \BibitemOpen
  \bibfield{author}{%
  \bibinfo {author} {\bibfnamefont{H.~M.}\ \bibnamefont{Wiseman}}\ and\
  \bibinfo {author} {\bibfnamefont{J.~A.}\ \bibnamefont{Vaccaro}},\ }%
  \bibfield{journal}{%
  \Doi{10.1103/PhysRevLett.91.097902}{\bibinfo {journal} {Phys. Rev. Lett.}}\
  }%
  \textbf{\bibinfo {volume} {91}},\ \bibinfo {pages} {097902} (\bibinfo {year}
  {2003})%
  \bibAnnoteFile{NoStop}{wiseman03a}%
\bibitem{kim02a}%
  \BibitemOpen
  \bibfield{author}{%
  \bibinfo {author} {\bibfnamefont{M.~S.}\ \bibnamefont{Kim}}, \bibinfo
  {author} {\bibfnamefont{W.}~\bibnamefont{Son}}, \bibinfo {author}
  {\bibfnamefont{V.}~\bibnamefont{Bu\v{z}ek}},\ and\ \bibinfo {author}
  {\bibfnamefont{P.~L.}\ \bibnamefont{Knight}},\ }%
  \bibfield{journal}{%
  \Doi{10.1103/PhysRevA.65.032323}{\bibinfo {journal} {Phys. Rev. A}}\ }%
  \textbf{\bibinfo {volume} {65}},\ \bibinfo {pages} {032323} (\bibinfo {year}
  {2002})%
  \bibAnnoteFile{NoStop}{kim02a}%
\bibitem{wolf03a}%
  \BibitemOpen
  \bibfield{author}{%
  \bibinfo {author} {\bibfnamefont{M.~M.}\ \bibnamefont{Wolf}}, \bibinfo
  {author} {\bibfnamefont{J.}~\bibnamefont{Eisert}},\ and\ \bibinfo {author}
  {\bibfnamefont{M.~B.}\ \bibnamefont{Plenio}},\ }%
  \bibfield{journal}{%
  \Doi{10.1103/PhysRevLett.90.047904}{\bibinfo {journal} {Phys. Rev. Lett.}}\
  }%
  \textbf{\bibinfo {volume} {90}},\ \bibinfo {pages} {047904} (\bibinfo {year}
  {2003})%
  \bibAnnoteFile{NoStop}{wolf03a}%
\bibitem{asboth05a}%
  \BibitemOpen
  \bibfield{author}{%
  \bibinfo {author} {\bibfnamefont{J.~K.}\ \bibnamefont{Asb\'oth}}, \bibinfo
  {author} {\bibfnamefont{J.}~\bibnamefont{Calsamiglia}},\ and\ \bibinfo
  {author} {\bibfnamefont{H.}~\bibnamefont{Ritsch}},\ }%
  \bibfield{journal}{%
  \Doi{10.1103/PhysRevLett.94.173602}{\bibinfo {journal} {Phys. Rev. Lett.}}\
  }%
  \textbf{\bibinfo {volume} {94}},\ \bibinfo {pages} {173602} (\bibinfo {year}
  {2005})%
  \bibAnnoteFile{NoStop}{asboth05a}%
\bibitem{jiang13a}%
  \BibitemOpen
  \bibfield{author}{%
  \bibinfo {author} {\bibfnamefont{Z.}~\bibnamefont{Jiang}}, \bibinfo {author}
  {\bibfnamefont{M.~D.}\ \bibnamefont{Lang}},\ and\ \bibinfo {author}
  {\bibfnamefont{C.~M.}\ \bibnamefont{Caves}},\ }%
  \bibfield{journal}{%
  \Doi{10.1103/PhysRevA.88.044301}{\bibinfo {journal} {Phys. Rev. A}}\ }%
  \textbf{\bibinfo {volume} {88}},\ \bibinfo {pages} {044301} (\bibinfo {year}
  {2013})%
  \bibAnnoteFile{NoStop}{jiang13a}%
\bibitem{Note2}%
  \BibitemOpen
  \bibinfo {note} {Note that one could also allow collective local unitaries
  (e.g., phase shifters), which would change the basis states, but not the
  entanglement structure in either picture.}%
  \bibAnnoteFile{Stop}{Note2}%
\bibitem{wootters98a}%
  \BibitemOpen
  \bibfield{author}{%
  \bibinfo {author} {\bibfnamefont{W.~K.}\ \bibnamefont{Wootters}},\ }%
  \bibfield{journal}{%
  \Doi{10.1103/PhysRevLett.80.2245}{\bibinfo {journal} {Phys. Rev. Lett.}}\ }%
  \textbf{\bibinfo {volume} {80}},\ \bibinfo {pages} {2245} (\bibinfo {year}
  {1998})%
  \bibAnnoteFile{NoStop}{wootters98a}%
\bibitem{wang03a}%
  \BibitemOpen
  \bibfield{author}{%
  \bibinfo {author} {\bibfnamefont{X.}~\bibnamefont{Wang}}\ and\ \bibinfo
  {author} {\bibfnamefont{B.~C.}\ \bibnamefont{Sanders}},\ }%
  \bibfield{journal}{%
  \Doi{10.1103/PhysRevA.68.012101}{\bibinfo {journal} {Phys. Rev. A}}\ }%
  \textbf{\bibinfo {volume} {68}},\ \bibinfo {pages} {012101} (\bibinfo {year}
  {2003})%
  \bibAnnoteFile{NoStop}{wang03a}%
\bibitem{wang04a}%
  \BibitemOpen
  \bibfield{author}{%
  \bibinfo {author} {\bibfnamefont{X.}~\bibnamefont{Wang}},\ }%
  \bibfield{journal}{%
  \Doi{http://dx.doi.org/10.1016/j.physleta.2004.08.019}{\bibinfo {journal}
  {Phys. Lett. A}}\ }%
  \textbf{\bibinfo {volume} {331}},\ \bibinfo {pages} {164 } (\bibinfo {year}
  {2004})%
  \bibAnnoteFile{NoStop}{wang04a}%
\bibitem{ulamorgikh01a}%
  \BibitemOpen
  \bibfield{author}{%
  \bibinfo {author} {\bibfnamefont{D.}~\bibnamefont{Ulam-Orgikh}}\ and\
  \bibinfo {author} {\bibfnamefont{M.}~\bibnamefont{Kitagawa}},\ }%
  \bibfield{journal}{%
  \Doi{10.1103/PhysRevA.64.052106}{\bibinfo {journal} {Phys. Rev. A}}\ }%
  \textbf{\bibinfo {volume} {64}},\ \bibinfo {pages} {052106} (\bibinfo {year}
  {2001})%
  \bibAnnoteFile{NoStop}{ulamorgikh01a}%
\bibitem{yin11a}%
  \BibitemOpen
  \bibfield{author}{%
  \bibinfo {author} {\bibfnamefont{X.}~\bibnamefont{Yin}}, \bibinfo {author}
  {\bibfnamefont{X.}~\bibnamefont{Wang}}, \bibinfo {author}
  {\bibfnamefont{J.}~\bibnamefont{Ma}},\ and\ \bibinfo {author}
  {\bibfnamefont{X.}~\bibnamefont{Wang}},\ }%
  \bibfield{journal}{%
  \Doi{10.1088/0953-4075/44/1/015501}{\bibinfo {journal} {J. Phys. B: At. Mol.
  Opt. Phys.}}\ }%
  \textbf{\bibinfo {volume} {44}},\ \bibinfo {pages} {015501} (\bibinfo {year}
  {2011})%
  \bibAnnoteFile{NoStop}{yin11a}%
\bibitem{coffman00a}%
  \BibitemOpen
  \bibfield{author}{%
  \bibinfo {author} {\bibfnamefont{V.}~\bibnamefont{Coffman}}, \bibinfo
  {author} {\bibfnamefont{J.}~\bibnamefont{Kundu}},\ and\ \bibinfo {author}
  {\bibfnamefont{W.~K.}\ \bibnamefont{Wootters}},\ }%
  \bibfield{journal}{%
  \Doi{10.1103/PhysRevA.61.052306}{\bibinfo {journal} {Phys. Rev. A}}\ }%
  \textbf{\bibinfo {volume} {61}},\ \bibinfo {pages} {052306} (\bibinfo {year}
  {2000})%
  \bibAnnoteFile{NoStop}{coffman00a}%
\bibitem{osborne06a}%
  \BibitemOpen
  \bibfield{author}{%
  \bibinfo {author} {\bibfnamefont{T.~J.}\ \bibnamefont{Osborne}}\ and\
  \bibinfo {author} {\bibfnamefont{F.}~\bibnamefont{Verstraete}},\ }%
  \bibfield{journal}{%
  \Doi{10.1103/PhysRevLett.96.220503}{\bibinfo {journal} {Phys. Rev. Lett.}}\
  }%
  \textbf{\bibinfo {volume} {96}},\ \bibinfo {pages} {220503} (\bibinfo {year}
  {2006})%
  \bibAnnoteFile{NoStop}{osborne06a}%
\bibitem{boixo08a}%
  \BibitemOpen
  \bibfield{author}{%
  \bibinfo {author} {\bibfnamefont{S.}~\bibnamefont{Boixo}}\ and\ \bibinfo
  {author} {\bibfnamefont{A.}~\bibnamefont{Monras}},\ }%
  \bibfield{journal}{%
  \Doi{10.1103/PhysRevLett.100.100503}{\bibinfo {journal} {Phys. Rev. Lett.}}\
  }%
  \textbf{\bibinfo {volume} {100}},\ \bibinfo {pages} {100503} (\bibinfo {year}
  {2008})%
  \bibAnnoteFile{NoStop}{boixo08a}%
\bibitem{simon02a}%
  \BibitemOpen
  \bibfield{author}{%
  \bibinfo {author} {\bibfnamefont{C.}~\bibnamefont{Simon}}\ and\ \bibinfo
  {author} {\bibfnamefont{J.}~\bibnamefont{Kempe}},\ }%
  \bibfield{journal}{%
  \Doi{10.1103/PhysRevA.65.052327}{\bibinfo {journal} {Phys. Rev. A}}\ }%
  \textbf{\bibinfo {volume} {65}},\ \bibinfo {pages} {052327} (\bibinfo {year}
  {2002})%
  \bibAnnoteFile{NoStop}{simon02a}%
\bibitem{stockton03a}%
  \BibitemOpen
  \bibfield{author}{%
  \bibinfo {author} {\bibfnamefont{J.~K.}\ \bibnamefont{Stockton}}, \bibinfo
  {author} {\bibfnamefont{J.~M.}\ \bibnamefont{Geremia}}, \bibinfo {author}
  {\bibfnamefont{A.~C.}\ \bibnamefont{Doherty}},\ and\ \bibinfo {author}
  {\bibfnamefont{H.}~\bibnamefont{Mabuchi}},\ }%
  \bibfield{journal}{%
  \Doi{10.1103/PhysRevA.67.022112}{\bibinfo {journal} {Phys. Rev. A}}\ }%
  \textbf{\bibinfo {volume} {67}},\ \bibinfo {pages} {022112} (\bibinfo {year}
  {2003})%
  \bibAnnoteFile{NoStop}{stockton03a}%
\bibitem{benatti12a}%
  \BibitemOpen
  \bibfield{author}{%
  \bibinfo {author} {\bibfnamefont{F.}~\bibnamefont{Benatti}}, \bibinfo
  {author} {\bibfnamefont{R.}~\bibnamefont{Floreanini}},\ and\ \bibinfo
  {author} {\bibfnamefont{U.}~\bibnamefont{Marzolino}},\ }%
  \bibfield{journal}{%
  \Doi{10.1103/PhysRevA.85.042329}{\bibinfo {journal} {Phys. Rev. A}}\ }%
  \textbf{\bibinfo {volume} {85}},\ \bibinfo {pages} {042329} (\bibinfo {year}
  {2012})%
  \bibAnnoteFile{NoStop}{benatti12a}%
\end{thebibliography}%

\end{document}